# A free boundary problem for fluid flow through porous media

C. Di Nucci

Dipartimento di Ingegneria Industriale e dell'Informazione e di Economia
Università degli Studi dell'Aquila (Italy)

carmine.dinucci@univaq.it

**Abstract**

This brief note addresses the free boundary problem arising from the steady two-dimensional seepage flow through a rectangular dam. The flow problem consists in finding the free boundary location, and the velocity and pressure fields. The problem solution is obtained by using an approximate model which provides the analytical expression for the free boundary profile. Within the range of validity of the model assumptions, the computational results show agreement with data in literature.

**1. Introduction**

This brief note deals with the steady two-dimensional seepage flow in a saturated porous medium. The flow problem consists in finding the free boundary location, the potential velocity field and the pressure field. From a mathematical point of view, the problem can be formulated as a free boundary problem for elliptic equations. Such problems arise in a wide variety of fluid mechanics applications [1].
For flow through porous media, the free boundary can be found by using numerical methods based either on finite element models [2] or on variational inequalities [3]. Approximate methods [4-5] are proposed to improve the performance of the Dupuit-Forchheimer theory [6].
For seepage flow in a rectangular dam, the free boundary is expressed by Polubarinova-Kochina (PK) equation [7]. This equation has been solved numerically in [8-9]. By using an approximate model, this brief note provides an analytical expression for the free boundary profile. The model generalizes the one proposed in [10] for steady axisymmetric potential flow. Within the range of validity of the model assumptions, the computational results show good agreement with PK equation.

**2. Flow problem**

Fig. 1 shows the definition sketch of the steady two-dimensional seepage flow through a rectangular dam. The dam is above a horizontal impermeable base and the porous medium is saturated, homogeneous and isotropic. The horizontal $x$-axis and the vertical $z$-axis form a Cartesian coordinate system. The free boundary is denoted by $h(x)$, the upstream flow depth by $h_1$, the downstream flow depth by $h_2$, the dam length by $L$, the seepage face by $h_{sf} = h(x = L) - h_2$. Neglecting the capillary fringe, the free boundary is the saturation line [6].
The flow problem consists in finding the location of the free boundary, the velocity field $\boldsymbol{v}(x,z)$ and the pressure field $p(x,z)$ in the unbounded flow domain $[0,L] \times [0,h(x)]$.
The momentum and continuity equations are expressed by Darcy law and solenoidality condition [11-12]:

$$\boldsymbol{v} = -k\nabla\left(z + \frac{p}{\rho g}\right) \tag{1}$$

$$\nabla \cdot \boldsymbol{v} = 0 \tag{2}$$

where $k$ is the constant hydraulic conductivity, $\rho$ the constant fluid density and $g$ the gravitational acceleration.

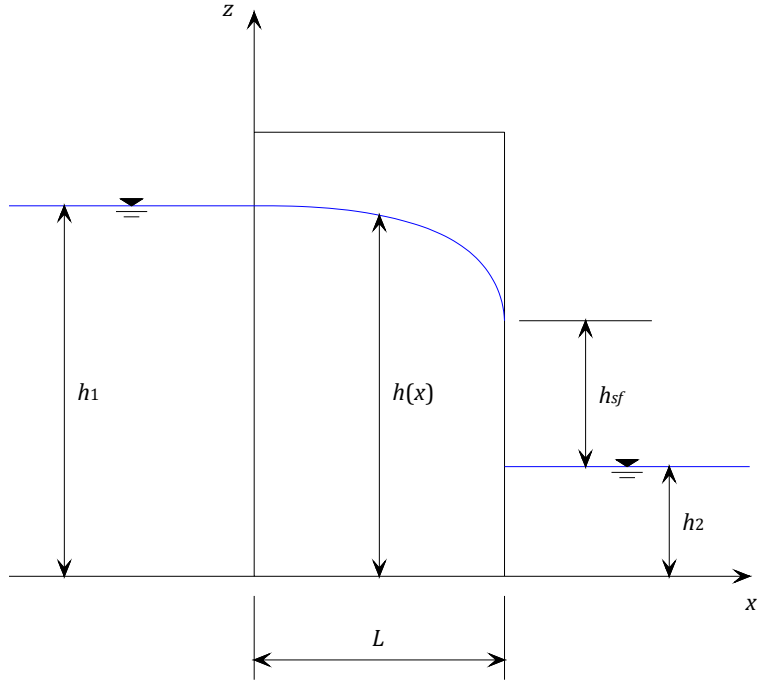

Fig. 1. Definition sketch of the flow problem.

Eq. (1) is consistent with irrotationalty condition:

$$\nabla \times \boldsymbol{v} = 0 \tag{3}$$

The boundary conditions are given as:

$$v_z(x, 0) = 0 \tag{4}$$

$$v_z(x, h) = v_x(x, h)\frac{\mathrm{d}h}{\mathrm{d}x} \tag{5}$$

$$p(x, h) = 0 \tag{6}$$

$$\frac{1}{\rho g} p(0, z) = h_1 - z \tag{7}$$

$$\frac{1}{\rho g} p(L, 0 \le z \le h_2) = h_2 - z \tag{8}$$

$$p(L, z > h_2) = 0 \tag{9}$$

where $v_x$ and $v_z$ are the horizontal and vertical velocity components, respectively.

Eqs. (4) and (5) are the kinematic boundary conditions at the base and at the free boundary; Eq. (6) is the dynamic boundary condition at the free boundary; Eq. (7) is the inflow boundary condition; Eqs. (8) and (9) are the outflow boundary conditions.

An analytical approximate solution can be obtained by using a procedure similar to that proposed in [10] for axisymmetric potential flow. By coupling approximate expressions for the velocity field to the momentum equation, the proposed model reduce the flow problem to one of finding the free boundary location and the flow rate $q$. In line with the adopted notation, the constant flow rate is given as:

$$q = \int_0^{h(x)} v_x \, \mathrm{d}x \tag{10}$$

As shown in Appendix, the velocity components can be expressed in an approximate form as:

$$v_x = \frac{q}{h} + \left[\frac{1}{2}\frac{q}{h^2}\frac{\mathrm{d}^2 h}{\mathrm{d}x^2} - \frac{q}{h^3}\left(\frac{\mathrm{d}h}{\mathrm{d}x}\right)^2\right] z^2 + \frac{1}{3}\frac{q}{h}\left(\frac{\mathrm{d}h}{\mathrm{d}x}\right)^2 - \frac{1}{6} q \frac{\mathrm{d}^2 h}{\mathrm{d}x^2} \tag{11}$$

$$v_z = \frac{q}{h^2}\frac{\mathrm{d}h}{\mathrm{d}x} z \tag{12}$$

With this setting, the pressure filed and the free boundary differential equation are given, respectively, as

$$\frac{p}{\rho g} = h - z - \frac{q}{2kh^2}\frac{\mathrm{d}h}{\mathrm{d}x}(z^2 - h^2) \tag{13}$$

$$\frac{\mathrm{d}^2 h}{\mathrm{d}x^2} = -\frac{1}{h}\left(\frac{\mathrm{d}h}{\mathrm{d}x}\right)^2 - \frac{3}{h} - \frac{3k}{q}\frac{\mathrm{d}h}{\mathrm{d}x} \tag{14}$$

Eq. (13) is obtained from the vertical momentum equation:

$$-k\frac{\partial}{\partial z}\left(z + \frac{p}{\rho g}\right) = \frac{q}{h^2}\frac{\mathrm{d}h}{\mathrm{d}x} z \tag{15}$$

subject to the boundary condition (6); Eq. (14) is the horizontal momentum equation:

$$-k\frac{\partial}{\partial x}\left(z + \frac{p}{\rho g}\right) = \frac{q}{h} + \left[\frac{1}{2}\frac{q}{h^2}\frac{\mathrm{d}^2 h}{\mathrm{d}x^2} - \frac{q}{h^3}\left(\frac{\mathrm{d}h}{\mathrm{d}x}\right)^2\right] z^2 + \frac{1}{3}\frac{q}{h}\left(\frac{\mathrm{d}h}{\mathrm{d}x}\right)^2 - \frac{1}{6} q \frac{\mathrm{d}^2 h}{\mathrm{d}x^2} \tag{16}$$

rewritten by using Eq. (13).

## 3. Results and discussion

For the problem under consideration, the flow rate is expressed as [6,13]:

$$q = k\frac{h_1^2 - h_2^2}{2L} \tag{17}$$

and the boundary conditions associated with Eq. (14) are given as [6]:

$$h(x = L) = h_1 \tag{18}$$

$$\frac{\mathrm{d}}{\mathrm{d}x} h(x = L) = 0 \tag{19}$$

The solution of Eq. (14), subject to Eqs. (18) and (19), is:

$$h(x) = \sqrt{h_1^2 - 2\frac{q}{k}x + \frac{2}{3}\frac{q^2}{k^2}\left[1 - \exp\left(-\frac{3k}{q}x\right)\right]} \quad (20)$$

Eqs. (17) and (20) solve the flow problem: once the flow rate and the free boundary are computed, the velocity and pressure fields are determined by Eqs. (11), (12) and (13).

Proceeding in analogy with [10], by imposing $v_x(x, h) > 0$ the range of validity of Eq. (19) can be estimated as:

$$\left(\frac{q}{kh}\right)^2 \left[\exp\left(-\frac{3k}{q}x\right) - 1\right] + 1 > 0 \quad (21)$$

As shown in Fig. 2, within this range, the free boundary profile deduced from Eq. (20) is in good agreement with numerical test data [8-9].

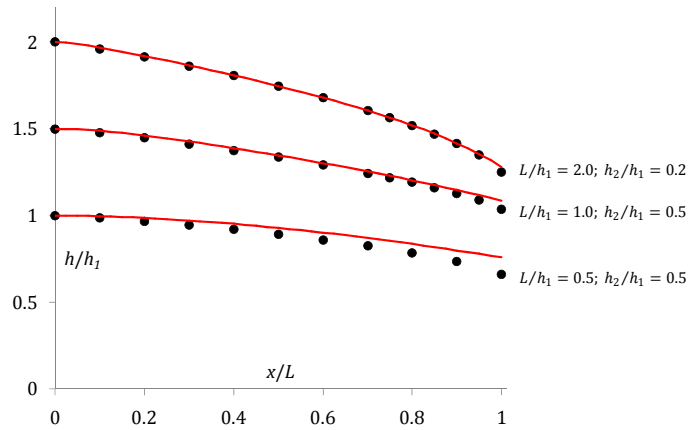

Fig. 2. Free boundary profiles: • numerical test data [8-9], — Eq. (20). The plots are shifted vertically by 0.5 units.

When condition (21) is violated, the free boundary has a slope and curvature such that Eqs. (11) and (12) are inappropriate to represent the velocity field. In this instance, it is necessary to increase the accuracy of the solution.

**Appendix**

The approximate velocity components can be found by using the following Picard iteration.
The iteration starts by putting in first approximation:

$$v_x = \frac{q}{h} \quad (A.1)$$

$$v_z = 0 \quad (A.2)$$

With this setting, Eqs. (3) and (4) are verified, while Eqs. (2) and (5) are fulfilled if and only if the following first-order term is small and can be neglected:

$$\frac{dh}{dx} \cong 0 \quad (A.3)$$

Eq. (A.3) defines the closure hypotheses of the first approximation.
The second approximation can be obtained by putting:

$$v_x = \frac{q}{h} \tag{A.4}$$

$$v_z = f_2(x, z) \tag{A.5}$$

where $f_2$ is a function a priori unknown. According to Eqs. (2) and (4), $f_2$ takes the form:

$$f_2(x, z) = \frac{q}{h^2}\frac{dh}{dx} z \tag{A.6}$$

With Eqs. (A.4), (A.5) and (A.6), Eq. (3) is verified if and only if the following second-order terms are small and can be neglected:

$$\left(\frac{dh}{dx}\right)^2 \cong 0, \quad \frac{d^2h}{dx^2} \cong 0 \tag{A.7}$$

These equations defines the closure hypotheses of the second approximation.
The third approximation can be deduced by setting:

$$v_x = \frac{q}{h} + f_3(x, z) \tag{A.8}$$

$$v_z = \frac{q}{h^2}\frac{dh}{dx} z \tag{A.9}$$

According to Eqs. (3) and (10), $f_3$ takes the form:

$$f_3 = \left[\frac{1}{2}\frac{q}{h^2}\frac{d^2h}{dx^2} - \frac{q}{h^3}\left(\frac{dh}{dx}\right)^2\right]z^2 + \frac{1}{3}\frac{q}{h}\left(\frac{dh}{dx}\right)^2 - \frac{1}{6}q\frac{d^2h}{dx^2} \tag{A.10}$$

With Eqs. (A.8), (A.9) and (A.10), Eq. (2) is verified if and only if the following third-order terms are small and can be neglected:

$$\left(\frac{dh}{dx}\right)^3 \cong 0, \quad \frac{d^2h}{dx^2}\frac{dh}{dx} \cong 0, \quad \frac{d^3h}{dx^3} \cong 0 \tag{A.11}$$

Eq. (A.11) defines the closure hypotheses of the third approximation.
This iterative procedure can be extended to any required degree of accuracy.
The approximate expressions found by this iterative procedure match with those obtained by using the power series expansion of the harmonic stream function [14-15] and the Picard iteration of Cauchy-Riemann equations [16].
The proposed procedure offers the advantage of providing the closure hypotheses of the approximate velocity field.